\documentclass[journal, 11pt, onecolumn]{IEEEtran}

\usepackage{physics}
\usepackage{graphicx}
\usepackage{enumerate}
\usepackage{amssymb, amsmath}
\usepackage{color}
\usepackage{hyperref}
\usepackage{caption}
\usepackage{csquotes}
\usepackage{natbib}
\usepackage[margin=1in]{geometry} 
\usepackage{titlesec}
\usepackage{subfigure}

\MakeOuterQuote{"}

\ifCLASSINFOpdf

\else

\fi

\begin{document}

\title{Full-waveform variational inference with full common-image gathers and diffusion network}

\author{Yunlin~Zeng\thanks{Yunlin Zeng is with the School of Physics, Georgia Institute of Technology},
Huseyin Tuna Erdinc\thanks{Huseyin Tuna Erdinc is with the School of Electrical and Computer Engineering, Georgia Institute of Technology}, 
Rafael Orozco\thanks{Rafael Orozco is with the School of Computational Science and Engineering, Georgia Institute of Technology},
Felix Herrmann\thanks{Felix Herrmann is with the School of Earth and Atmospheric Sciences, School of Computational Science and Engineering, and School of Electrical and Computer Engineering, Georgia Institute of Technology.}}

\maketitle

\begin{abstract}
Accurate seismic imaging and velocity estimation are essential for subsurface characterization. Conventional inversion techniques, such as full-waveform inversion, remain computationally expensive and sensitive to initial velocity models. To address these challenges, we propose a simulation-based inference framework with conditional elucidated diffusion models for posterior velocity-model sampling. Our approach incorporates both horizontal and vertical subsurface offset common-image gathers to capture a broader range of reflector geometries, including gently dipping structures and steep dipping layers. Additionally, we introduce the background-velocity model as an input condition to enhance generalization across varying geological settings. We evaluate our method on the SEAM dataset, which features complex salt geometries, using a patch-based training approach. Experimental results demonstrate that adding the background-velocity model as an additional conditioning variable significantly enhances performance, improving SSIM from $0.717$ to $0.733$ and reducing RMSE from $0.381\,$km/s to $0.274\,$km/s. Furthermore, uncertainty quantification analysis shows that our proposed approach yields better-calibrated uncertainty estimates, reducing uncertainty calibration error from $6.68\,$km/s to $3.91\,$km/s. These results show robust amortized seismic inversion with uncertainty quantification.

\end{abstract}



\IEEEpeerreviewmaketitle

\section{Introduction}
introduction
Accurate seismic imaging and reliable velocity-model estimation are critical for understanding subsurface geological structures. Conventional full-waveform inversion (FWI) is a powerful method for subsurface imaging, but it often suffers from strong non-linearity, high computational costs, and sensitivity to the initial velocity model \citep{Virieux2009, Warner2013}. Recently, simulation-based inference (SBI) methods \citep{Cranmer2020}—which leverage pairs of simulated seismic observations and their corresponding ground-truth velocity models $\left\{ \mathbf{x}^{(i)}, \mathbf{y}^{(i)} \right\}_{i = 1}^N$—have emerged as promising alternatives to traditional inversion techniques. These methods utilize machine learning models to amortize the inversion process, enabling efficient inference of subsurface velocity from new observational data, while also providing systematic Bayesian uncertainty quantification \citep{zhang2023bayesian}.

One notable SBI framework is Wavefield Inversion using conditional normalizing flows (WISE), which samples posterior velocity models conditioned on seismic common-image gathers (CIGs) \citep{yin2023wise}. By leveraging CIGs, the WISE framework efficiently extracts physics-informed summary statistics from seismic shot data and enables inversion with uncertainty quantification. However, horizontal CIGs alone may be insufficient for imaging complex geological settings—particularly those involving steeply dipping or vertical reflectors—due to geometric limitations associated with horizontal offsets \citep{Shan&Biondi2008, vanleeuwen2015EAGEafs}.

We propose an SBI approach that incorporates both horizontal and vertical subsurface offset image gathers to include full information of migrated wavefields. The addition of vertical offsets extends the imaging condition to better capture steeply dipping reflectors by correlating wavefields at vertically shifted depths \citep{Biondi2006, Sava&Vasconcelos2011, vanleeuwen2015EAGEafs}. Furthermore, instead of using conditional normalizing flows as in \citet{yin2023wise}, we adopt score-based diffusion generative networks. Recent studies have demonstrated that score-based methods \citep{ho2020denoising, song2020score, karras2022} offer superior sample quality compared to CNFs, particularly in capturing multimodal distributions and generating realistic posterior samples for complex seismic inverse problems \citep{orozco2024machinelearningenabledvelocitymodel}.

In this study, we extend the SBI framework by integrating both horizontal and vertical CIGs into a score-based diffusion generative network \citep{karras2022}. Additionally, we propose using the background-velocity model as an extra conditioning variable alongside the CIGs. We argue that including the background model not only broadens the amortization capacity of the network—enabling generalization across diverse geological settings—but also improves sample quality \citep{elsemuller2024sensitivityaware}. To validate our approach, we present inference results on the SEAM dataset, which features complex salt dome structures \citep{fehler2011model}.

\section{Method}

Our proposed method employs a two-stage process involving: (1) generation of extended CIGs (both horizontal and vertical), and (2) posterior velocity-model inference using the score-based network conditioned on these extended gathers.

\subsection{Extended Imaging Conditions: Horizontal and Vertical CIGs}

Seismic imaging typically involves correlating the forward-propagated source wavefield $\mathbf{u}$ and backward-propagated receiver wavefield $\mathbf{v}$. The zero-offset imaging condition, while common, does not capture velocity-model inaccuracies or steeply dipping reflectors effectively \citep{Nolet1986}. Extended imaging conditions address this by introducing non-zero spatial offsets into the wavefield correlation, resulting in extended-image gathers that directly encode velocity-related information. When the velocity model is correct, energy focuses at zero offset; deviations from zero indicate velocity errors \citep{Sava&Fomel2005, Shan&Biondi2008, biondi+symes2004}.

Horizontal CIGs correlate wavefields horizontally displaced around the imaging point and are defined as:
\begin{equation}
I_x(x,z,h_x) = \iint \mathbf{u}(s_x,x - h_x,z,t)\, \mathbf{v}(s_x,x + h_x,z,t)\, dt\, ds_x,
\end{equation}
where $h_x$ is horizontal offset, $t$ is propagation time, and $s_x$ is the horizontal source location \citep{Biondi2006}. Horizontal offsets capture reflection curvature indicative of velocity inaccuracies and work best for reflectors with gentle to moderate dips \citep{Rickett2002OffsetAA}.

In contrast, vertical CIGs correlate wavefields at vertically displaced imaging points:
\begin{equation}
I_z(x,z,h_z) = \iint \mathbf{u}(s_x,x,z - h_z,t)\, \mathbf{v}(s_x,x,z + h_z,t)\, dt\, ds_x,
\end{equation}
where $h_z$ is the vertical offset \citep{Shan&Biondi2008}. Vertical subsurface offsets work best at imaging steeply dipping or vertical reflectors, which require impractically large horizontal offsets if only horizontal CIGs are used \citep{Biondi&SHan2002}. By combining both horizontal and vertical offsets, our method generates comprehensive subsurface images capable of accurately representing complex geological structures and robustly informing velocity updates.

\subsection{Conditional Diffusion for Seismic Inference}

Diffusion generative networks are neural density estimation algorithms that approximate the data distribution by learning the score function, defined as the gradient of the log-likelihood $\nabla_{\mathbf{x}}\log p(\mathbf{x})$ \citep{song2020score}. These networks operate by progressively perturbing data samples into a Gaussian distribution through a stochastic process and then learning to reverse this transformation to generate new samples from the target distribution.

A key component in diffusion networks is the noise schedule, denoted by $\sigma(t)$, which governs the gradual corruption of data over time. As the time variable $t$ increases, the data distribution becomes increasingly smoothed (the PDF becomes more noisy), eventually converging to a standard Gaussian. The network is trained to estimate the score function $\nabla_{\mathbf{x}} \log p(\mathbf{x}, \sigma(t))$ at each time step, enabling it to reconstruct the original distribution during inference.

Once trained, it can perform likelihood estimation and generate new samples by numerically solving a reverse-time stochastic differential equation. In our seismic inversion work, we employ elucidated diffusion models (EDM), which provide a well-conditioned and computationally efficient framework for both training and sampling \citep{karras2022}. EDM simplify the standard score-matching objective and introduce an optimized noise schedule, leading to improved sample quality and reduced computational cost.

\subsection{Conditional Diffusion for Seismic Imaging}

In seismic inversion, we are interested in sampling from the posterior distribution $p(\mathbf{x} \mid \mathbf{y})$, where $\mathbf{x}$ represents the velocity model and $\mathbf{y}$ denotes the CIGs obtained from migrated seismic shot data. To achieve this, we modify the standard diffusion framework to learn the conditional score function $\nabla_{\mathbf{x}} \log p(\mathbf{x} \mid \mathbf{y})$.

Following prior work on conditional diffusion \citep{batzolis2021, orozco2024machinelearningenabledvelocitymodel}, we incorporate the observation $\mathbf{y}$ as an additional conditioning variable in the score-matching objective:
\begin{equation}
\hat{\theta} = \underset{\boldsymbol{\theta}}{\operatorname{arg min}} \, \mathbb{E}_{\mathbf{y}\sim p(\mathbf{y}\mid\mathbf{x})} \mathbb{E}_{\mathbf{n}\sim \mathcal{N}(0,t^2 I)} \| s_{\boldsymbol{\theta}}(\mathbf{x}+\mathbf{n}, \mathbf{y}, t)-\mathbf{x} \|^2_2,
\end{equation}
where $s_{\boldsymbol{\theta}}(\mathbf{x}, \mathbf{y}, t)$ represents the learned score function, and $\mathbf{n} \sim \mathcal{N}(0, t^2 I)$ denotes the Gaussian noise perturbation.

To train the model, we construct a dataset $\mathcal{D} = \{(\mathbf{x}^{(i)},\mathbf{y}^{(i)})\}_{i=0}^{N}$, where each sample pair consists of a ground-truth velocity model and the corresponding CIGs obtained through seismic migration. The training process involves minimizing the following loss:
\begin{equation}
\hat{\theta} = \underset{\boldsymbol{\theta}}{\operatorname{arg min}} \sum_{i=0}^{N} \mathbb{E}_{\mathbf{n}\sim \mathcal{N}(0,t^2 I)}\| s_{\boldsymbol{\theta}}(\mathbf{x}^{(i)}+\mathbf{n},\mathbf{y}^{(i)},t)-\mathbf{x}^{(i)} \|^2_2.
\end{equation}

After training, posterior samples can be generated by conditioning the model on new observations $\mathbf{y}_{\text{obs}}$ and solving the reverse SDE, analogous to standard unconditional diffusion models.

\subsection{Incorporating Vertical and Horizontal CIGs into Diffusion Model}

Our approach conditions the diffusion model on both horizontal and vertical offset CIGs to enhance posterior velocity sampling, which introduces depth shifts in wavefield correlations and captures reflector geometries that horizontal offsets alone cannot resolve \citep{Shan&Biondi2008, vanleeuwen2015EAGEafs}.

In our model, we use $50$ channels of horizontal offsets and $50$ channels of vertical offsets. In addition, we include a background-velocity model as a separate channel, which serves as a reference condition. Including the background as an additional input channel expands the amortization scope of our network and generalizes across different velocity backgrounds during inference.

\section{Experiment Setup}

We conduct experiments using 2D slices extracted from the 3D SEAM model \citep{fehler2011model}, which features complex salt geometries characteristic of the Gulf of Mexico. To create a robust training dataset while ensuring diversity, we select a continuous range of crosslines for training and enforce a minimum separation of $2$ km between the nearest training and test slices to reduce redundancy.

For seismic data simulation, we generate narrow-offset 2D seismic lines with a spatial grid of $1744 \times 512$ and a grid spacing of $20$ m. The shot interval is set to $1000$ m to optimize computational efficiency, while receivers are densely sampled at $20$ m intervals with a maximum offset of $6$ km. The recorded shot gathers have a total duration of $9$ s, and Gaussian noise is added to achieve a signal-to-noise ratio of $25$ dB within the source frequency band. Each velocity model contains $32$ shot records and assumes a constant density with spatially varying velocity.

To create initial migration-velocity models, we preprocess the ground-truth velocity models by removing salt bodies and smoothing the result using a Gaussian kernel with a grid size of $25$. The subsurface-offset migration is performed with $50$ evenly distributed offsets between $-2000$ m and $2000$ m. After migration, subsets of the full seismic lines are extracted to form training pairs, using sliding windows of $512 \times 512$ grids, resulting in $770$ training samples.

We adopt a patch-based training strategy for our conditional diffusion model \citep{orozco2024machinelearningenabledvelocitymodel}, training on smaller velocity patches while performing inference on full-grid resolution ($512 \times 1744$). This is made possible by the convolutional architecture of the network, which enables flexible evaluation on larger spatial domains without requiring retraining. As a result, the model scales effectively to high-resolution domains. Training each model requires approximately 12 GPU hours per iteration.

\section{Results}

\subsection{Quality of Posterior Samples}

\begin{figure}
\centering
\includegraphics[width=0.45\linewidth]{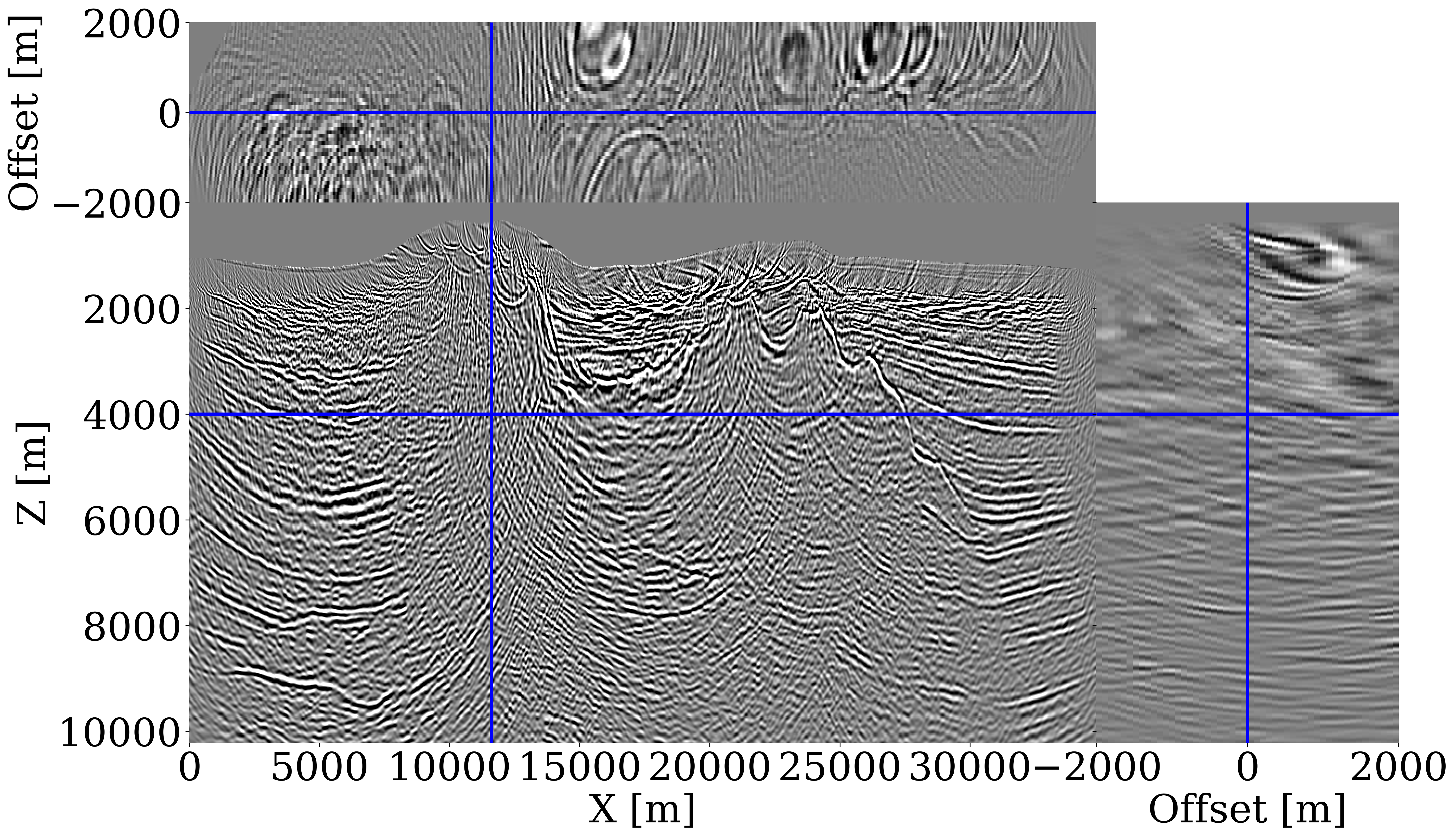}
\includegraphics[width=0.45\linewidth]{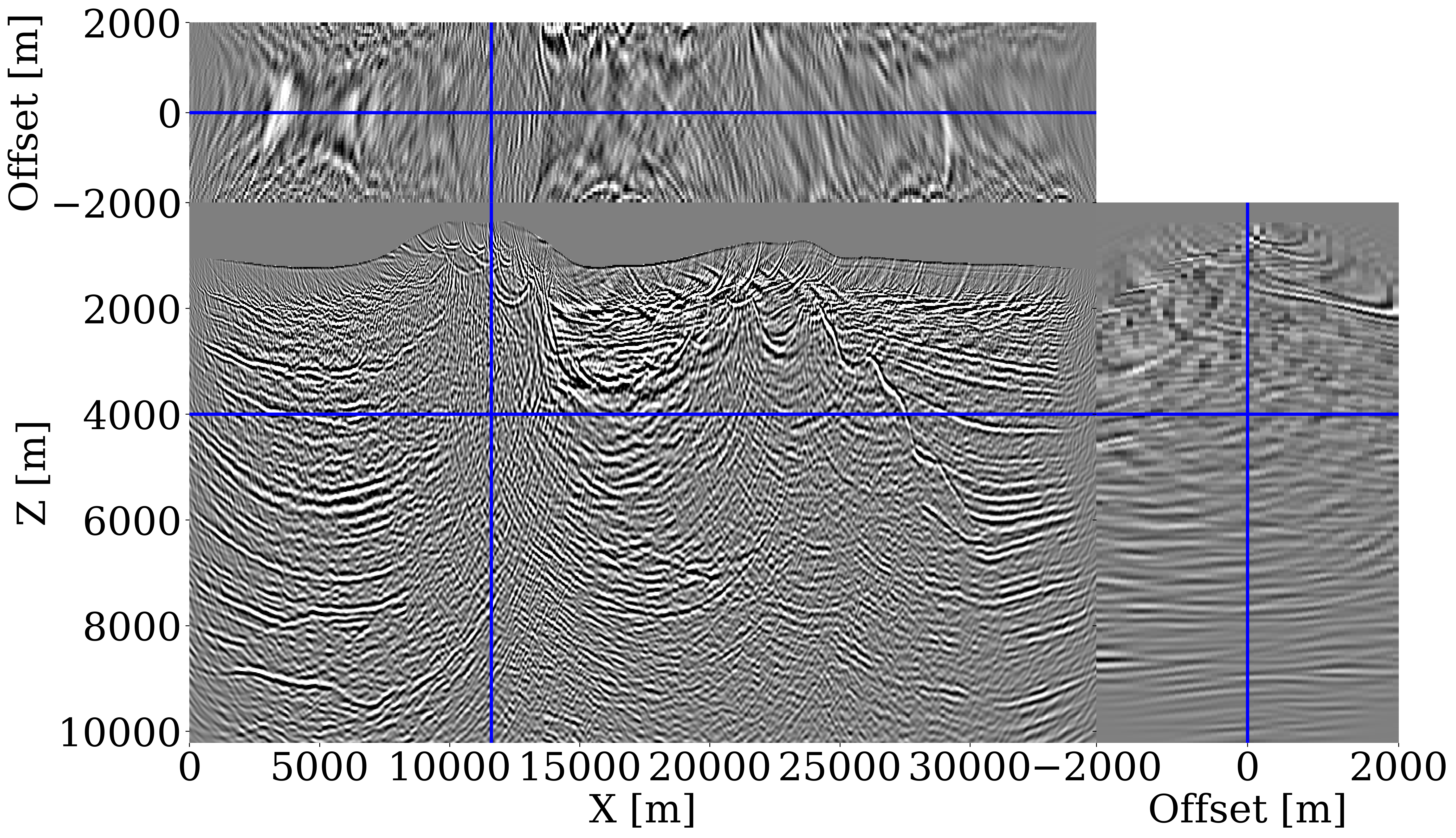}
\caption{Common-image gather (CIGs) plots. \textit{left}: Horizontal subsurface offset CIGs. \textit{right:} Vertical subsurface offset CIGs.}
\label{fig-cig}
\end{figure}

\begin{figure}
\centering
\includegraphics[width=0.7\linewidth]{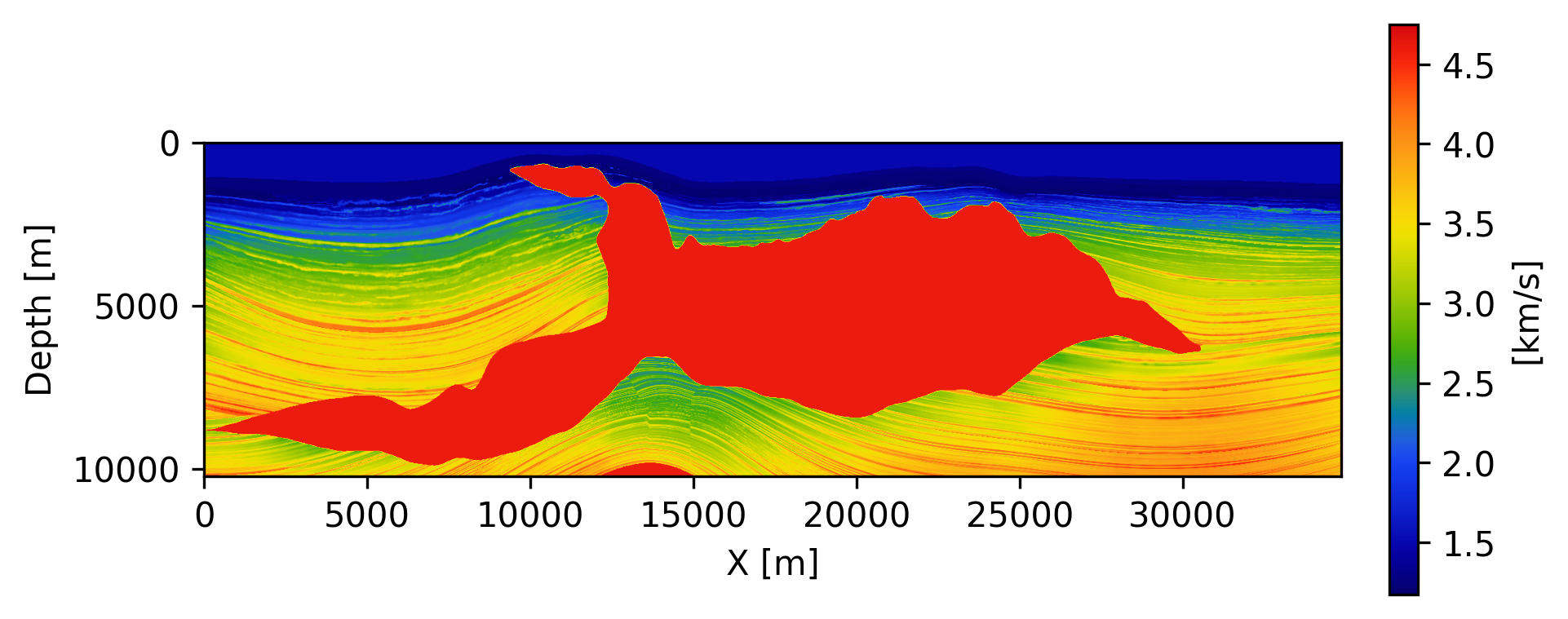}
\caption{One ground truth velocity model in the test set.}
\label{fig-x-gt}
\end{figure}

\begin{figure}
\centering
\includegraphics[width=0.7\linewidth]{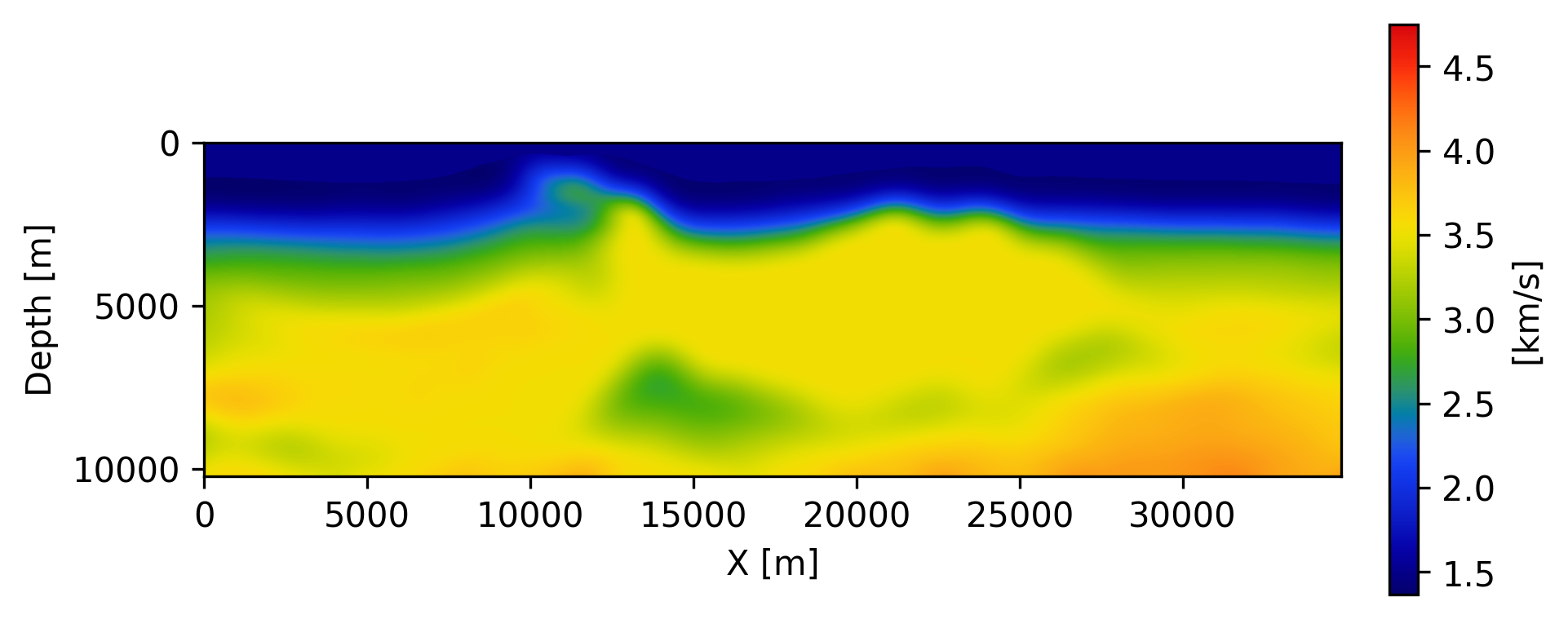}
\caption{The migration-velocity model.}
\label{fig-v0}
\end{figure}

\begin{figure}
\centering
\includegraphics[width=0.7\linewidth]{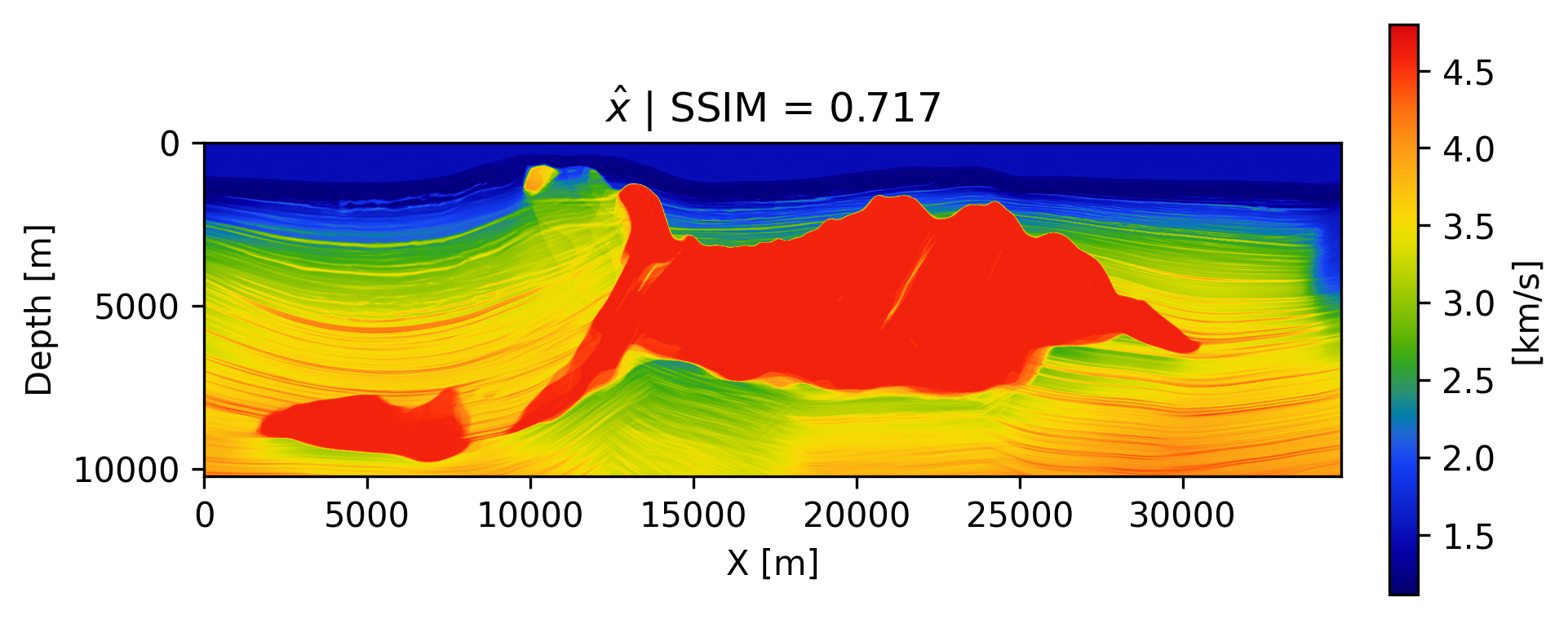}
\caption{Posterior mean velocity model inferred using horizontal and vertical CIGs without incorporating the background as an additional input.}
\label{fig-x-mean}
\end{figure}

\begin{figure}
\centering
\includegraphics[width=0.7\linewidth]{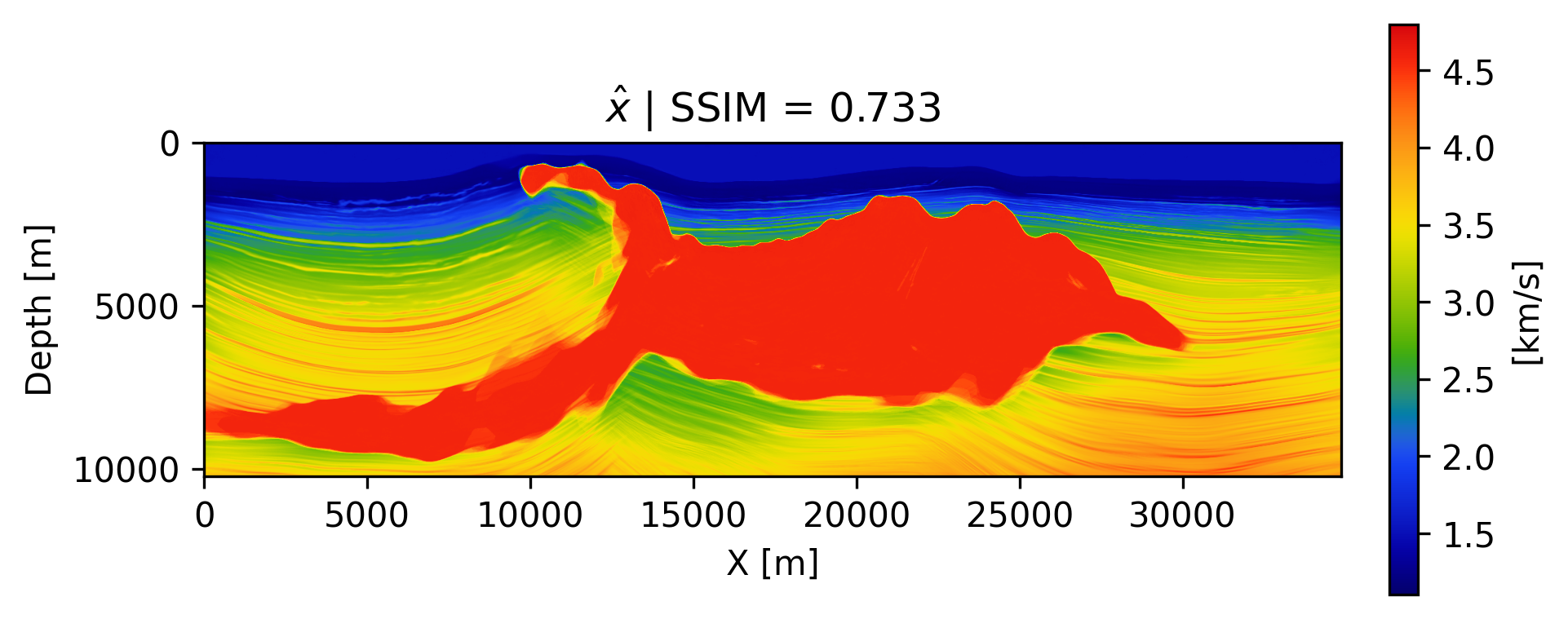}
\caption{Posterior mean velocity model inferred using horizontal and vertical CIGs with incorporating the background as an additional input.}
\label{fig-x-mean-w-bkg}
\end{figure}

\begin{figure}
\centering
\includegraphics[width=0.7\linewidth]{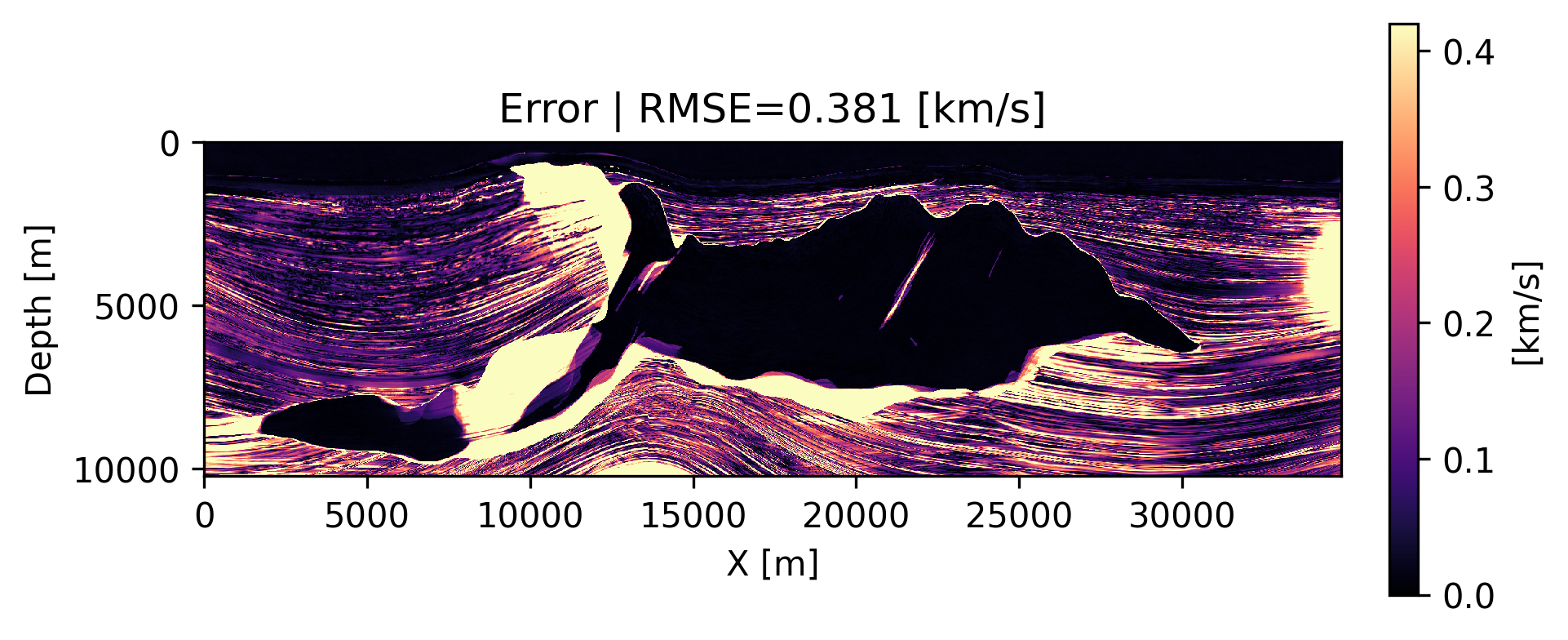}
\caption{Absolute error between the posterior mean in Fig.~\ref{fig-x-mean} and the ground truth velocity model.}
\label{fig-error}
\end{figure}

\begin{figure}
\centering
\includegraphics[width=0.7\linewidth]{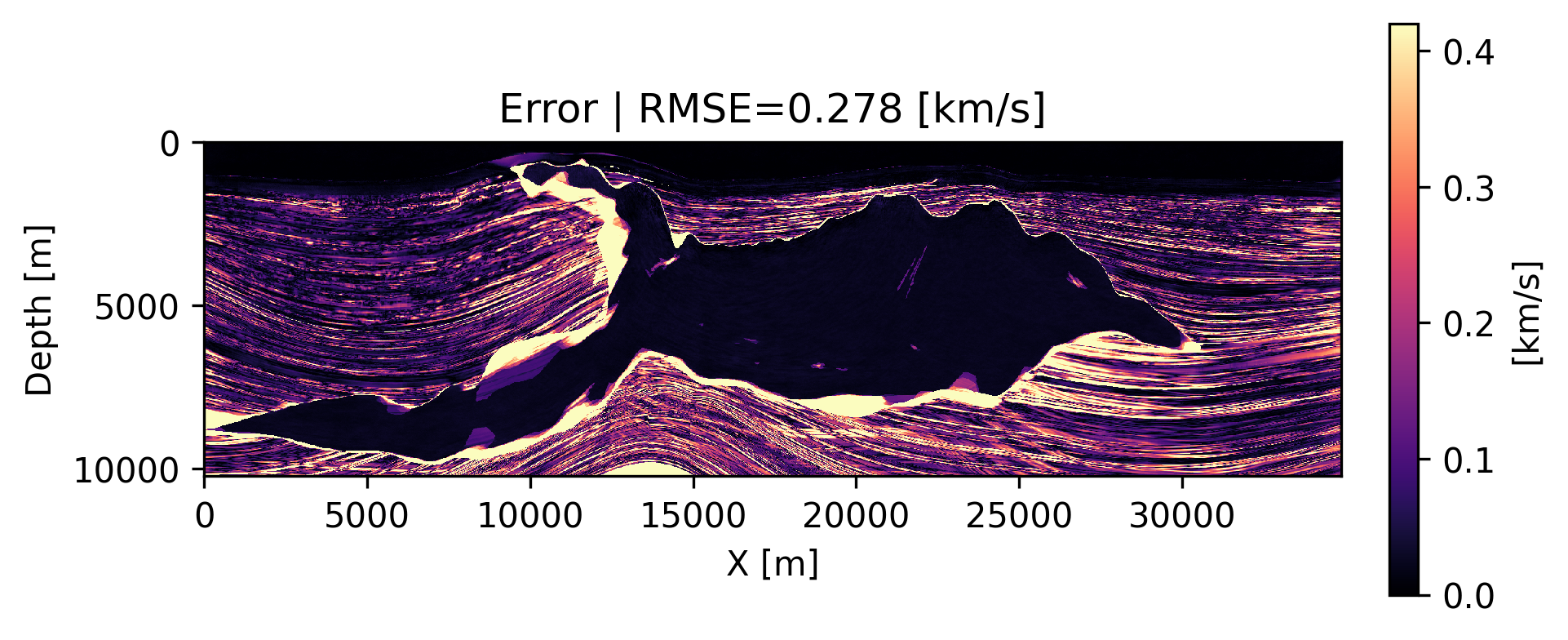}
\caption{Absolute error between the posterior mean in Fig.~\ref{fig-x-mean-w-bkg}. and the ground truth velocity model.}
\label{fig-error-w-bkg}
\end{figure}

Using EDM conditioned on both horizontal (Fig.~\ref{fig-cig} left) and vertical (Fig.~\ref{fig-cig} right) CIGs migrated from the background in Fig.~\ref{fig-v0}, we further evaluate the impact of incorporating a background-velocity model as an additional input channel. Compared to the baseline model that conditions only on horizontal and vertical CIGs (Fig.~\ref{fig-x-mean} and Fig.~\ref{fig-error}), the inclusion of the background-velocity model significantly improves reconstruction quality (Fig.~\ref{fig-x-mean-w-bkg} and Fig.~\ref{fig-error-w-bkg}). Specifically, the structural similarity index measure (SSIM) increases from $0.717$ to $0.733$, and the root-mean-square error (RMSE) decreases from $0.381\,$km/s to $0.274\,$km/s. These metrics are computed by comparing the posterior mean velocity models to the ground truth velocity model shown in Fig.~\ref{fig-x-gt}.

The improvement in SSIM and RMSE shows the benefit of using background-velocity information to help the network’s ability to generalize across different subsurface conditions and resolve velocity ambiguities in seismic inversion. The additional background input leads to sharper geological layer reconstructions, particularly in the deeper regions beneath the salt structure, and the bottom left part of the salt.

\section{Uncertainty Quantification}

Beyond sample quality, accurately quantifying uncertainty is critical for reliable seismic inversion. We use calibration of uncertainty to evaluate the quality of uncertainty quantification.

\subsection{Calibration of Uncertainty}

The calibration of uncertainty evaluates how well the predicted uncertainties align with observed errors \citep{guo2017calibration, laves2020well}. Ideally, the predicted uncertainty ($\hat{\sigma}$) should match the observed error magnitude exactly. Formally, perfect calibration satisfies:

\begin{equation}
\mathbb{E}{\mathbf{x}_\text{gt}, \bar{\mathbf{y}}}\left[|\mathbf{x}_\text{gt} - \hat{\mathbf{x}}| \mid \hat{\sigma} = \sigma\right] = \sigma, \quad \forall \sigma \geq 0.
\end{equation}

To quantify calibration, predicted uncertainties ($\hat{\sigma}$) and observed errors ($|\mathbf{x}_\text{gt} - \hat{\mathbf{x}}|$) are binned based on uncertainty magnitude. For each bin $B_k$, we compute average uncertainty in bin $B_k$

\begin{equation}
\text{UQ}(B_k) = \frac{1}{|B_k|}\sum_{i\in B_k}\hat{\sigma}_i,
\end{equation}

and average error in bin $B_k$

\begin{equation}
\text{Error}(B_k) = \frac{1}{|B_k|}\sum_{i\in B_k}|\mathbf{x}_{\text{gt}, i}-\hat{\mathbf{x}}_i|.
\end{equation}

The averages of uncertainty and error in each bin are plotted against each other, with ideal calibration represented by points falling on the $45^\circ$ line. The deviation from this line is quantified by the Uncertainty Calibration Error (UCE), defined as the area between the calibration curve and the $45^\circ$ line:

\begin{equation}
\text{UCE} = \sum_{k} |B_k|\cdot|\text{UQ}(B_k)-\text{Error}(B_k)|.
\end{equation}

\begin{figure}
\centering
\includegraphics[width=0.6\linewidth]{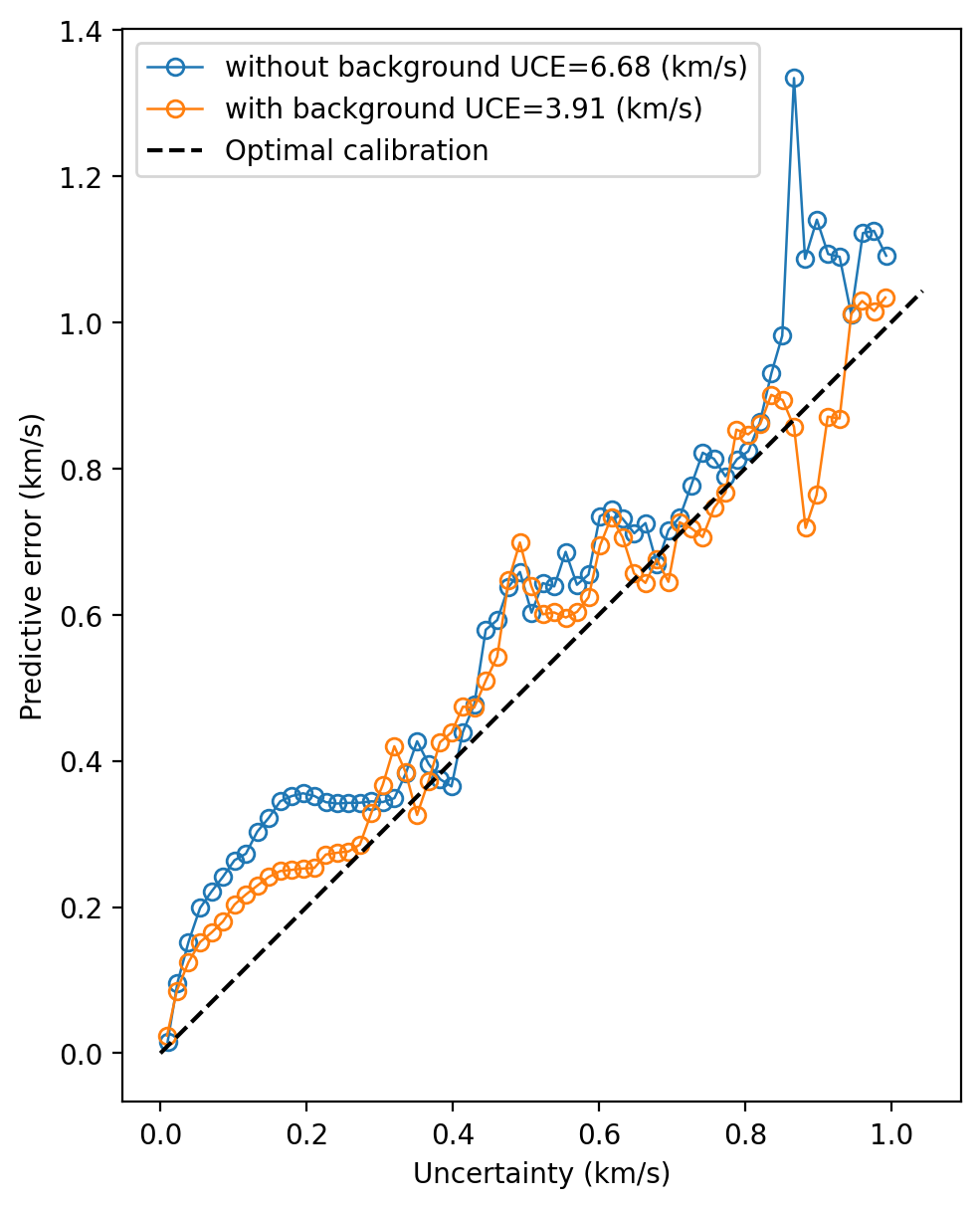}
\caption{Comparison of uncertainty calibration between the diffusion network without (blue) and with (orange) a background channel. The orange curve aligns more closely with the optimal $45^\circ$ calibration line, and the UCE is reduced.}
\label{fig-calibration}
\end{figure}

Ideally, uncertainty calibration curves should align with the 45° line for a perfect agreement. In Fig.~\ref{fig-calibration}, we compare the calibration performance of diffusion models conditioned on different CIG with and without background input. Both curves exhibit a positive correlation between predicted uncertainty and actual error. However, incorporating the background as an additional input improves calibration, as seen in the orange curve aligning more closely with the optimal line than the baseline (blue curve). To quantify this, we compute the uncertainty calibration error (UCE), which measures the deviation between predicted uncertainty and observed error across all bins. Adding the background channel reduces UCE from $6.68\,$km/s to $3.91\,$km/s.

\section{Conclusion and Discussion}

In this work, we introduce a conditional score-based generative model for seismic velocity inversion, using the EDM network. By incorporating horizontal and vertical CIGs, our method achieves decent inversion performance. Additionally, we demonstrate that including the background-velocity model as a conditioning variable significantly improves inference performance. Specifically, the SSIM improves from $0.717$ to $0.733$, the RMSE decreases from $0.381\,$km/s to $0.274\,$km/s, and the UCE is reduced from $6.68\,$km/s to $3.91\,$km/s, indicating improved uncertainty estimation reliability.

Future work will explore extending this framework to 3D seismic inversion. Additionally, we aim to further enhance the diffusion network by training separate networks conditioned individually on horizontal and vertical CIGs and subsequently aligning these representations in the latent space. This latent-space alignment, inspired by twin-network and multi-view learning approaches \citep{Chen2020, Tian2020}, has the potential to more effectively extract information from both types of gathers. Overall, our findings demonstrate that conditional score-based generative models, when properly conditioned on extended imaging gathers and background velocity, offer a powerful alternative to traditional seismic inversion methods.


\section*{Acknowledgment}

This research was carried out with the support of Georgia Research Alliance and partners of the ML4Seismic Center. 

During the preparation of this work, the authors used ChatGPT to refine sentence structures and improve the readability of the manuscript. After using this service, the authors reviewed and edited the content as needed and take full responsibility for the content of the publication.

\bibliographystyle{plainnat}
\bibliography{refs}



\end{document}